\documentclass[aps,nofootinbib,preprint,tightenlines]{revtex4}

\usepackage{graphicx}
\usepackage{amsmath}
\usepackage{bm} 
\usepackage{color}
\usepackage{amssymb}

\newcommand{\beq}{\begin{equation}}
\newcommand{\eeq}{\end{equation}}
\newcommand{\bqa}{\begin{eqnarray}}
\newcommand{\eqa}{\end{eqnarray}}
\newcommand{\del}{\partial}

\newcommand{\gsim}{\hspace*{0.2em}\raisebox{0.5ex}{$>$}
     \hspace{-0.8em}\raisebox{-0.3em}{$\sim$}\hspace*{0.2em}}
\newcommand{\lsim}{\hspace*{0.2em}\raisebox{0.5ex}{$<$}
     \hspace{-0.8em}\raisebox{-0.3em}{$\sim$}\hspace*{0.2em}}
\def\mqo2{{\!\!\!}}
\renewcommand{\vec}{\mathbf}

\begin{document}


\preprint{HISKP-TH-10-29}
\title{Resonant three-body physics in two spatial dimensions}
\author{K.\ Helfrich}

\author{H.-W.\ Hammer}
\affiliation{Helmholtz-Institut f\"ur Strahlen- und Kernphysik (Theorie)\\
and Bethe Center for Theoretical Physics,
 Universit\"at Bonn, 53115 Bonn, Germany}

\date{\today}

\begin{abstract}
We discuss the three-body properties of identical bosons
exhibiting large scattering length in two spatial dimensions. 
Within an effective field theory for resonant interactions,
we calculate the leading non-universal corrections from the
two-body effective range to bound-state and scattering observables.
In particular, we compute the three-body binding energies, the
boson-dimer scattering properties, and the three-body
recombination rate for finite energies. We find significant 
effective range effects for three-body observables
in the vicinity of the unitary limit.
The implications of this result for future experiments are
briefly discussed.
\end{abstract}

\maketitle

\section{Introduction}
\label{sec:intro}
In the last few years, ultracold quantum gases have become 
a versatile tool to investigate few- and many-body phenomena
in strongly interacting quantum systems.
With the help of Feshbach resonances, the atomic interaction
strength can be changed at will. This allows, for example, for the 
investigation of the BCS-BEC crossover in Fermi gases, the creation of
molecules, the observation of the Efimov
effect and of other universal phenomena~(see,
e.g.,~\cite{Bloch,Ferlaino} and references therein). The possibility
of using optical lattices makes them also interesting for the 
simulation of condensed matter problems such as the Hubbard 
model~\cite{Esslinger}. 
Special trap geometries allow for the creation of
lower-dimensional systems. They can, for example, help to understand
high-temperature superconductivity which is a two-dimensional ($2d$) problem.
Moreover, $2d$ systems are interesting on their own
since their behavior can be qualitatively different from three
dimensions.

Here, we concentrate on few-body phenomena in ultracold
atomic gases with large scattering length.
Such phenomena are of great interest because they are 
insensitive to the details of the interaction at short distances.
They can be described in an expansion around the unitary limit. This
limit refers to an idealized system where the range of the
interaction is taken to zero and the scattering length $a$ is infinite. To
leading order in this expansion, the low-energy observables are
universal. They are 
determined by the scattering length $a$ of the particles alone.
The leading non-universal corrections are due to the effective
range of the interaction. In the current paper, we focus on these 
corrections. Since there is no Efimov effect~\cite{Efimov70} in two 
dimensions~\cite{Bruch,Nielsen99}, three-body interactions are suppressed
and enter only at higher orders.

The definition of the scattering length $a$ in $2d$
is ambiguous since $\cot\delta$ diverges logarithmically as the 
wave number $k$ approaches zero and different conventions are used in
the literature. We follow the conventions of 
Verhaar et al.~\cite{Verhaar},
in which the effective range expansion of the scattering
phase shift is given by
\begin{equation}
\cot\delta(k)=\frac{2}{\pi}\left\{
 \gamma_E+\ln\left( \frac{ka}{2} \right)\right\}
   + \frac{r^2}{2\pi} k^2 + {\mathcal O}(k^4) \,,
\label{eq:ere1}
\end{equation}
where $\gamma_E \simeq 0.577216$ is Euler's constant.
Note that the scattering length in two dimensions is always positive.

\begin{figure}
	\centerline{\includegraphics*[width=5cm]{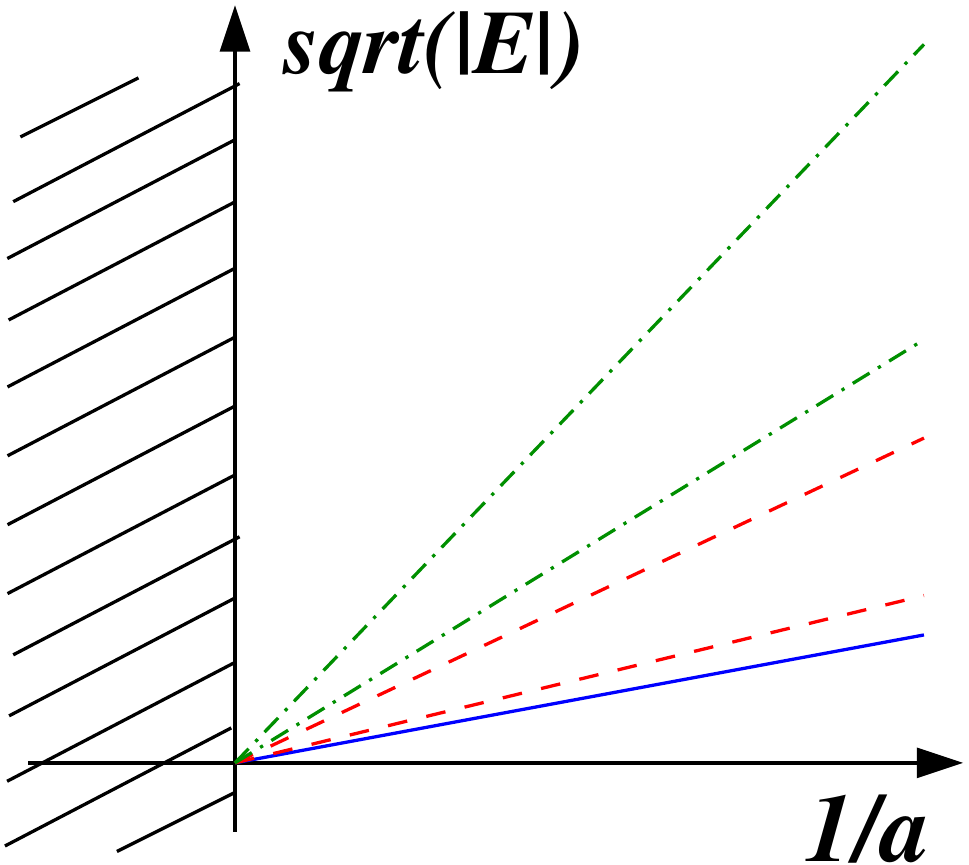}}
	\caption{(Color online) Spectrum of universal two-, three-, and four-body states
                 in two spatial dimensions as a function of the inverse 
                 scattering length $1/a$.}
\label{fig:spectrum}
\end{figure}

In the limit $a\gg|r|$,
the binding energy of the shallow dimer is universal,
\begin{eqnarray}
\label{eq:Edimer}
E_2 = 4 e^{-2 \gamma_E} \frac{\hbar^2}{ma^2} + {\mathcal O}({r^2 /a^2})   
\,. 
\end{eqnarray}
The binding energies of three- and four-body states in this limit 
have been calculated by various groups and are also 
universal~\cite{Bruch,Nielsen99,Son,PHM,Kagan2006,Karta}.
Since there is no other parameter in the problem, 
the energies must be multiples of the dimer energy.
There are two three-body bound states which were first calculated
by Bruch and Tjon \cite{Bruch}. Their binding energies are~\cite{Son}
\begin{eqnarray}
E_3^{(1)} & = & 1.2704091(1) \, E_2 \,\quad\mbox{ and }\quad
E_3^{(0)} =  16.522688(1) \, E_2 \,,
\label{eq:3bdy2D}
\end{eqnarray}
where the number in parentheses indicates the numerical 
error in the last quoted digit.
The first calculation of the four-body bound states in $2d$ was carried
out by Platter et al.~\cite{PHM}.
They found also two universal bound states with binding energies
\begin{eqnarray}
E_4^{(1)} & = & 25.5(1) \, E_2 \,
\quad\mbox{ and }\quad
E_4^{(0)} =  197.3(1) \, E_2 \,.
\label{eq:4bdy2D}
\end{eqnarray}
These results were later confirmed in Ref.~\cite{Kagan2006}.
In Fig.~\ref{fig:spectrum}, we illustrate the scattering length dependence
of this spectrum.
The universal few-body states do not cross
the continuum threshold $E=0$ for any finite value of the scattering length.
In contrast to three dimensions, the $2d$ universal states can 
therefore not be 
observed as zero-energy resonances in few-body recombination. 

For large values of $N\gg 1$, one can derive the universal 
properties of shallow $N$-boson ground states close to the
unitary limit~\cite{Son}. 
In particular, the binding energy
$E_N$ of the $N$-boson ground state increases geometrically with $N$:
\begin{equation}
  \label{BNratio}
  \frac{E_{N+1}}{E_N} \approx 8.567, \qquad N\gg 1 \,.
\end{equation}
Thus, the separation energy for one particle 
is approximately 88\% of the
total binding energy.  This is in contrast to most other physical
systems, where the ratio of the single-particle separation energy 
to the total binding energy decreases to zero as the number of particles 
increases. The numbers $E_3^{(0)}/E_2=16.5$ and 
$E_4^{(0)}/E_3^{(0)}=11.9$ obtained from the exact 3-body and
4-body results in Eqs.~(\ref{eq:3bdy2D}) and~(\ref{eq:4bdy2D}) appear to
be converging toward the universal prediction for large $N$
in Eq.~(\ref{BNratio}). In Ref.~\cite{Lee:2005nm}, $E_N$ was explicitly
calculated up to $N=10$ in lattice effective field theory and 
found to be consistent with Eq.~(\ref{BNratio}).
In any real physical system, however, the relation~(\ref{BNratio}) can only be 
valid up to some maximum value of $N$ determined by the range of the 
underlying interaction. When the states become compact enough that 
short-distance properties are probed, the 
binding energy  will no longer be universal.
In particular, for Lennard-Jones potentials and realistic
Helium-Helium potentials, effective range effects
can be quite large for three and more particles
and the universal limit is approached only slowly~\cite{Blume2005}.

In experiments with cold atoms in a trap, the quasi-$2d$ limit can be reached 
by special trap geometries.
The influence of a trapping potential on ultracold gases in 
this limit was extensively studied by Petrov and
collaborators~\cite{Petrov,Petrov2,Petrov3}. Although these works are
mainly concerned with many-body effects in two-dimensional systems,
they have applications for few-body aspects as well.

In this paper, we concentrate on strictly 
two-dimensional systems of bosons neglecting any trapping effects. We 
calculate three-body observables close to unitarity in the framework
of an effective field theory for large scattering 
length. We are especially interested in the leading non-universal
corrections due to effective range effects. They enter at 
next-to-leading order in the effective field theory.
Such effects must be under control for the experimental
observation of universal phenomena in $2d$.
In particular, we calculate the  leading non-universal
corrections to the three-body binding energies,
the boson-dimer scattering phase shift and effective range
parameters, and  the three-body recombination 
rate for finite energy.

\section{Method}
\label{sec:method}
In this section, we briefly review the derivation of the three-body
equations for $d=2$ in effective field theory. 
(See, e.g., Refs.~\cite{Son,Braaten:2004rn} for more details.)
For convenience, we set $\hbar=1$ in the following equations. 
We include a boson field $\Psi$ and an auxiliary dimer field $d$
in the Lagrangian. Since we include effective range effects, 
the dimer field is dynamical:
\beq
\label{eq:lagrangian}
{\cal L}=\Psi^\dagger\left(i\del_t+\frac{\nabla^2}{2
  m}\right)\Psi+d^\dagger\left(\eta\left(i\del_t+\frac{\nabla^2}{4
  m}\right)+\Delta\right)d-\frac{g}{4}(d^\dagger\Psi^2+\Psi^{\dagger 2}d)
+\ldots\, ,
\eeq
where the dots indicate higher order terms,
$m$ is the mass of the particles, $\eta=\pm 1$, and $\Delta$ 
and $g$ denote the bare coupling constants.
The sign $\eta$ can be used to tune the sign of the effective range
term. Negative $\eta$ leads to positive values of the 
effective range $r^2$. In this case, the dimer kinetic term has a
negative sign and the dimer field is a ghost. 
We will come back to this issue below.
Note that three-body interactions enter only at higher orders
and are not considered in this work.

The $2d$ effective range expansion, Eq.~(\ref{eq:ere1}),
can also be written in terms of the binding wave number 
$\kappa=\sqrt{m\,E_2}$:
 \bqa
 \label{eq:ere2d}
 \cot\delta(k)
 &=&\frac{2}{\pi}\ln\left(\frac{k}{\kappa}\right)
+\frac{r^2}{2\pi}(\kappa^2+k^2)+ {\mathcal O}(k^4) \, .
 \eqa
We can deduce the dependence of the binding wave number on
the scattering length and the effective range
from Eqs.~(\ref{eq:ere1}) and~(\ref{eq:ere2d}), 
\bqa
\label{eq:kappad}
\kappa&=&-\frac{i}{r}\sqrt{2\, W\left(-2e^{-2\gamma_E}\frac{r^2}{a^2}\right)}
\approx\frac{2e^{-\gamma_E}}{a}\sqrt{1+2e^{-2\gamma_E}\frac{r^2}{a^2}}\, ,
\eqa
where the signs are chosen such that $\kappa>0$. The function $W$ is the
product logarithm or Lambert $W$-function. It is defined as the solution
to $z=we^w$, namely $W(z)=w$. 
In the limit $r^2 \rightarrow 0$ the expression for $E_2$ 
reduces to Eq.~(\ref{eq:Edimer}) and $\kappa\approx 1.1229/a$.

\begin{figure}
	\centerline{\includegraphics*[width=8cm]{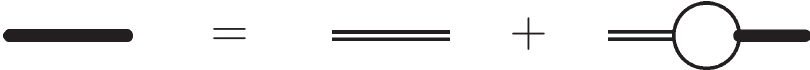}}
	\caption{Integral equation for the full dimer propagator
          (thick solid line). The bare dimer propagator and the boson 
          propagator are indicated by double and single lines, respectively.}
\label{fig:bubbles}
\end{figure}

The Lagrangian in Eq.~(\ref{eq:lagrangian}) implies the
following Feynman rules: The propagator for a boson with energy $k_0$ 
and wave number $\vec{k}$ is given by $i/(k_0-k^2/(2m)-i\epsilon)$,
where $k=|\vec{k}|$. The bare dimer propagator is
$i/(\eta(k_0-k^2/(4m))+\Delta)$ and the 
boson-dimer vertex coupling is given by $-i g/2$. 
Because the scattering length is large, boson loops are not
suppressed and the bare propagator has to be 
dressed by boson bubbles to all orders. The full
dimer propagator can be obtained by solving the integral equation 
in Fig.~\ref{fig:bubbles}. This leads to the expression
\beq
\label{eq:prop1} 
iD(p_0,p)=-i\frac{32\pi}{mg^2}\left\{\ln\left[
\frac{p^2/4-mp_0-i\epsilon}{\kappa^2}\right]+\frac{r^2}{2}
\left(\kappa^2+mp_0-p^2/4\right)\right\}^{-1}\, ,
\eeq
where we have already matched $g$ and $\Delta$ to 
the effective range expansion, Eq.~(\ref{eq:ere2d}).
The wave function renormalization constant is given by the residue of the 
bound state pole in the propagator~(\ref{eq:prop1}):
\beq
Z=\frac{32\pi}{m^2g^2}\frac{2\kappa^2}{2-\kappa^2 r^2}\, .
\eeq

\begin{figure}
	\centerline{\includegraphics*[width=8cm]{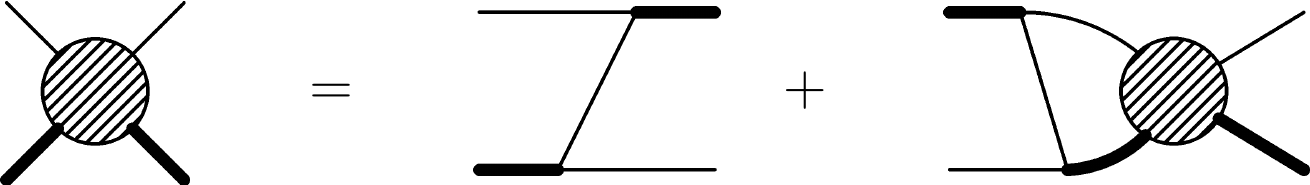}}
	\caption{Integral equation for the boson-dimer scattering amplitude.
                 The boson (full dimer) propagators are indicated by the solid
                 (thick solid) lines. The external lines are amputated.}
\label{fig:STM2d}
\end{figure}
The boson-dimer scattering amplitude is given by the integral equation 
in Fig.~\ref{fig:STM2d}, which iterates the 
one-boson exchange to all orders. 
Using the Feynman rules from above and projecting onto $S$-waves, we
obtain~\cite{Braaten:2004rn,Son}:
\bqa
\label{eq:STM2}
T(p,k;E)&=&\frac{16\pi}{m}\frac{\kappa^2}{2-\kappa^2
  r^2}\frac{1}{\sqrt{(p^2+k^2-mE)^2-p^2k^2}} \nonumber \\
&+&4\int_0^\infty\frac{dq\, 
  q\, T(q,k;E)}{\sqrt{(p^2+q^2-mE)^2-p^2q^2}}\nonumber\\
&\times&\left(\ln\left[\frac{\frac{3}{4}q^2-mE-i\epsilon}{\kappa^2}\right]
+\frac{r^2}{2}\left(\kappa^2+mE-\frac{3}{4}q^2\right)\right)^{-1}\, ,
\eqa
where $k$ ($p$) are the relative wave numbers of the incoming
(outgoing) boson and dimer in the center-of-mass system
and $E$ is the total energy.
The amplitude $T(p,k;E)$ has simple poles at negative energies 
corresponding to three-body bound states. A more
general discussion of the analytic properties of few-body scattering 
amplitudes in $2d$ is given in Ref.~\cite{Adhikari1992}.

The three-body binding energies are most easily obtained from solving
the homogeneous version of Eq.~(\ref{eq:STM2}) for negative energies
$E=-E_3$:
\bqa
\label{eq:STM2bs}
B(p;E_3)&=&
4\int_0^\infty\frac{dq\, 
  q\, B(q;E_3)}{\sqrt{(p^2+q^2+mE_3)^2-p^2q^2}}\nonumber\\
&\times&\left(\ln\left[\frac{\frac{3}{4}q^2+mE_3}{\kappa^2}\right]
+\frac{r^2}{2}\left(\kappa^2-mE_3-\frac{3}{4}q^2\right)\right)^{-1}\, .
\eqa
In Eqs.~(\ref{eq:STM2}) and (\ref{eq:STM2bs}) the effective
range $r^2$ is included nonperturbatively in the denominator
of the dimer propagator (\ref{eq:prop1}). Both equations therefore
contain some higher-order effective range effects but still
correspond to next-to-leading order in the effective field theory 
expansion. At the next higher order, where terms proportional to
$(r^2)^2$ enter, there are also contributions from
the $k^4$ term in the effective range expansion~(\ref{eq:ere1}),
(\ref{eq:ere2d}) which are not included here.\footnote{Note 
that in three dimensions the $k^4$ term 
enters one order higher and the corresponding equation would be valid
to next-to-next-to-leading order. 
The difference is due to the form of the effective
range expansion in two dimensions.}

The integral equations~(\ref{eq:STM2}) and~(\ref{eq:STM2bs}) can be solved
in a straightforward way for negative effective range ($\eta=1$). 
For positive effective range ($\eta=-1$),
an unphysical deep bound state pole appears in the dimer 
propagator~(\ref{eq:prop1}). As the effective range is increased,
this pole moves to lower energies. Its appearance
is related to a violation of the Wigner
causality bound which constrains the value of the
effective range $r^2$ for short-ranged, energy-independent interactions.
For a detailed discussion of this bound in general dimension $d$, 
see Refs.~\cite{Hammer:2009zh,Hammer:2010fw}.
This deep pole appears when we circumvent the Wigner bound by introducing 
a ghost dimer field ($\eta=-1$). It limits the energy range where
our approach is applicable.
Identifying the position space cutoff in~\cite{Hammer:2009zh}
with $1/\Lambda$, the Wigner bound translates to
\beq
r^2 \leq \frac{2}{\Lambda^2}\left\{ \left[ \ln(\Lambda a)+1/2\right]^2
+1/4\right\}\,,
\label{eq:wigbo}
\eeq
where $\Lambda$ is an ultraviolet cutoff on the integration wave numbers
in Eqs.~(\ref{eq:STM2}) and~(\ref{eq:STM2bs}).
In the limit $\Lambda \to \infty$, the constraint becomes $r^2 \leq 0\,.$
There are at least two strategies to deal with this problem:
\begin{enumerate}
\item 
\label{s:1}
Expand the full dimer propagator~(\ref{eq:prop1}) 
to linear order in $r^2$ and treat the range perturbatively. This 
removes the deep pole and includes all terms to next-to-leading order.
\item 
\label{s:2}
Keep an explicit wave number cutoff $\Lambda$ in equations~(\ref{eq:STM2}) 
and~(\ref{eq:STM2bs}) such that Eq.~(\ref{eq:wigbo})
is satisfied. The unphysical pole then has no effect on low-energy
observables.
\end{enumerate}
Both strategies are applicable for wave numbers $|k^2 r^2|\ll 1$.
In the following, we make use of Eqs.~(\ref{eq:STM2}) and~(\ref{eq:STM2bs})
and use strategy~\ref{s:2} to calculate three-body  
observables. A brief description of the perturbative 
treatment is given in Appendix~\ref{app:pert}.

\section{Three-body observables}
\label{sec:Results}
In this section, we present our results for the leading non-universal
corrections to the three-boson binding energies, the boson-dimer 
scattering phase shifts and effective range parameters, and  
the three-boson recombination rate for finite energy. Since we are
mainly interested in applications to cold atoms, we will refer to
the bosons as atoms in the remainder of the paper.
\subsection{Three-body binding energies}
We start with the effective range corrections to the three-body 
binding energies in Eq.~(\ref{eq:3bdy2D}). For $r^2\kappa^2<0$, the energies
can straightforwardly be obtained by solving Eq.~(\ref{eq:STM2bs}). This 
is similar to the three-dimensional case investigated in~\cite{Petrov4}.
For $r^2\kappa^2>0$, we have to keep track of the Wigner bound. 
We use an explicit wave number cutoff $\Lambda$ and vary $\Lambda$ 
from 1/5 to 4/5 of the maximum value determined by the position of the 
unphysical pole
in Eq.~(\ref{eq:STM2bs}). This value agrees within a factor of two
with the maximum value given by Eq.~(\ref{eq:wigbo}). The dependence
of the three-body energies on the cutoff is monotonic. The smallest
cutoff results in the smallest energy value whereas the largest
cutoff results in the largest energy. This cutoff 
variation allows us to check whether the calculation is converged 
with respect to the cutoff and gives an error estimate for our results.
We note that
one still has to be careful about possible artefacts from the iteration
of range terms. In the $3d$ case, it was shown that the ultraviolet behavior
of the integral-equation kernel is already modified for momenta well below
$1/r$~\cite{Platter:06}.

\begin{figure}
	\centerline{\includegraphics*[width=12cm]{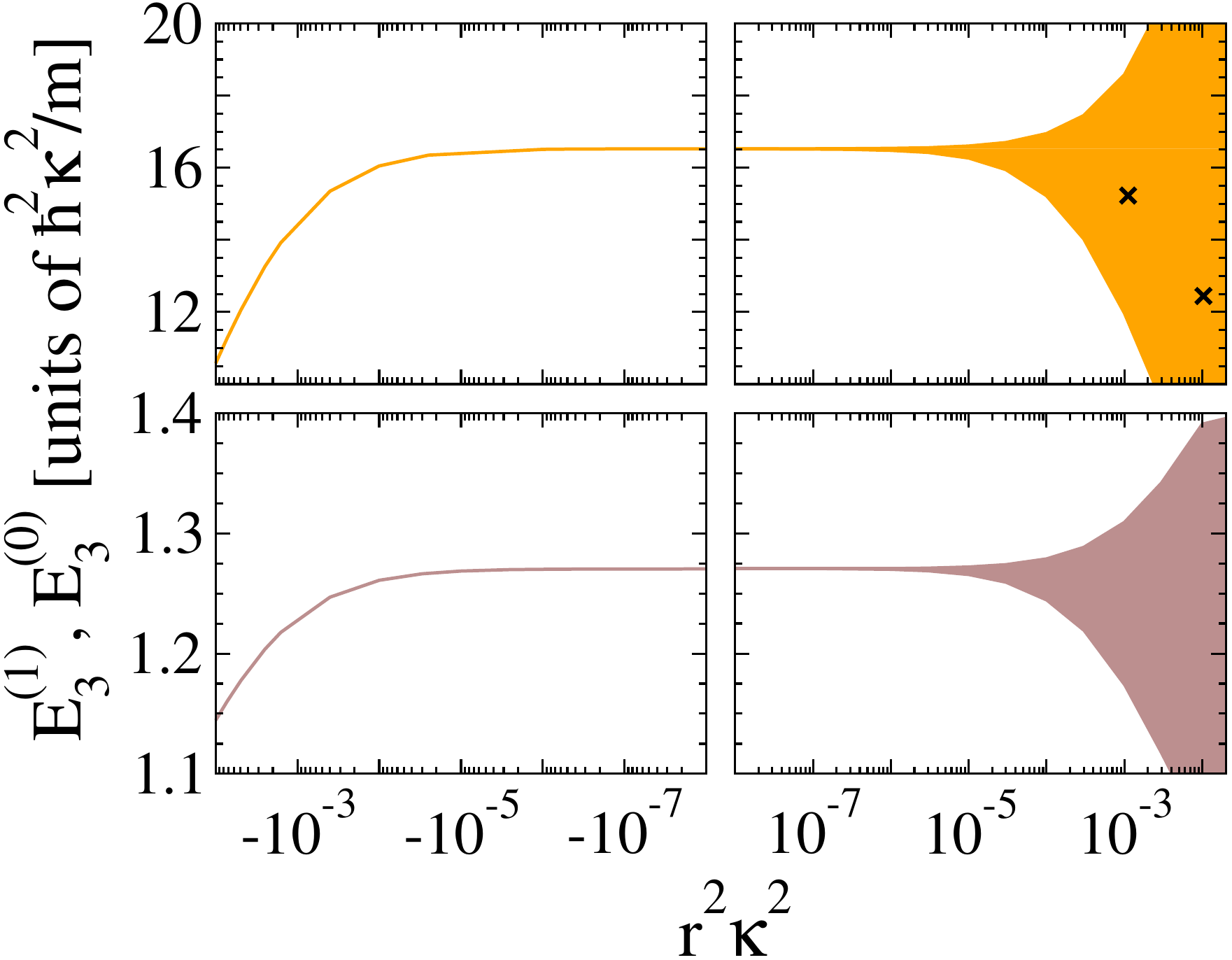}}
	\caption{(Color online) Three-body binding energies 
          $E_3^{(1)}$ and $E_3^{(0)}$ in units of $\hbar^2\kappa^2/m$ vs.\ the
          two-body effective range $r^2 \kappa^2$. 
          The shaded bands are derived with the help of cutoff
          variation as described in the text and provide an error estimate.
          The crosses are Monte Carlo results for the 
            modified KORONA potential
            from \cite{Blume2005,Blume:pc}.}
\label{fig:bindenergy}
\end{figure}

Our results for the three-body binding energies $E_3^{(1)}$ and 
$E_3^{(0)}$ as a function of the effective range, $r^2\kappa^2$,
are summarized in Fig.~\ref{fig:bindenergy}.
For negative effective range $r^2\kappa^2$, both three-body states become 
less bound as $|r^2\kappa^2|$ is increased. This behavior is quantitatively 
similar to the dimer state, cf.~Eq.~(\ref{eq:kappad}).
The binding energies are very 
sensitive to the effective range. For $r^2\kappa^2= -0.01$, 
we find the values $E_3^{(1)}=1.145(1)\,E_2$ for the excited state
and $E_3^{(0)}=10.578(1)\,E_2$ for the ground state. For this
rather small effective range,
the ground state energy has already shifted by about 30\% while the 
excited state energy is shifted 
by about 10\%. This sensitivity is partially
related to the special nature of the 
effective range term in $2d$ which has units of [length]$^2$. 
Taking the square root, the leading range correction of $10-30$\% for 
$|r \kappa | = 0.1$ looks more natural.
Our calculation including the leading 
non-universal corrections suggests that the three-body states
eventually cross the atom-dimer threshold as the effective range is 
made more negative. For the excited state this happens around 
$r^2\kappa^2 \approx -0.4$, but higher order corrections are expected
to be important. If this behavior holds true and the effective
range could be varied in experiment, the three-body states in
$2d$ might be observable through zero energy scattering resonances
similar to Efimov states in $3d$.
For positive values of the effective range, 
the central value of our
error band also corresponds to a less strongly bound system but
we can not make a definite prediction. Once the effective range effects 
become appreciable, the errors in our calculation become too large.
However, the Monte Carlo results for the modified KORONA potential 
from \cite{Blume2005,Blume:pc} (see the crosses in Fig.~\ref{fig:bindenergy})
are in good agreement with our band. They also
give a diminishing energy for larger positive effective range.
Note that we only show the two points from \cite{Blume2005} closest
to the unitary limit. All other data points are outside the range of
$r^2 \kappa^2$ displayed in Fig.~\ref{fig:bindenergy}.

For effective ranges close to zero, the binding energies depend 
linearly on $r^2\kappa^2$. We can determine the coefficient of
the leading term numerically to about 15\% accuracy. 
For $r^2\kappa^2<0$, we find:
\bqa
\nonumber
E_3^{(0)}/E_2&=&16.522688(1)+28000(5000)\, r^2\kappa^2+{\cal O}(r^4\kappa^4)
\,,\\
E_3^{(1)}/E_2&=&1.2704091(1)+540(80)\, r^2\kappa^2+{\cal O}(r^4\kappa^4)\,.
\label{eq:E3pert}
\eqa
For $r^2\kappa^2>0$, the coefficient of $r^2\kappa^2$ can not 
be extracted from our calculation. While the values of $E_3^{(0)}$
and $E_3^{(1)}$ 
are very insensitive to the cutoff variation at small positive $r^2\kappa^2$,
even the sign of the slope is not well determined.
We note that $r^2 \kappa^2$ can be quite small in $2d$ systems even if
the effective range is substantially larger that the range of the 
interaction (see, e.g., the explicit example of a circular well 
potential given in Sec.~6.1 of Ref.~\cite{Hammer:2010fw}).
The large coefficients in Eq.~(\ref{eq:E3pert}), however, cause
significant effective range corrections already for $|r^2 \kappa^2|$
of order $10^{-4}$ to $10^{-3}$.

\subsection{Atom-dimer scattering}
\begin{figure}
	\centerline{\includegraphics*[width=11cm]{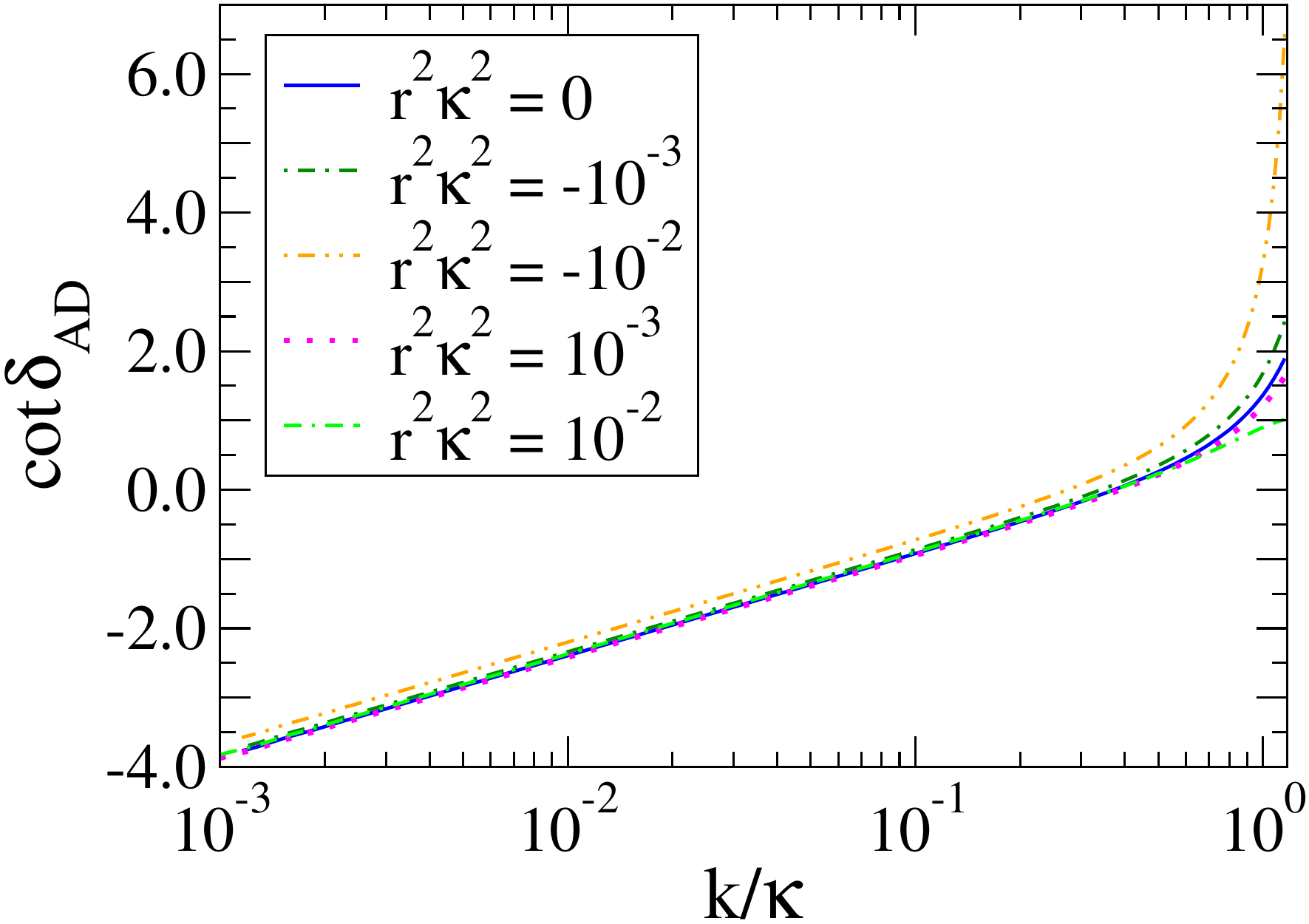}\qquad}
	\caption{(Color online) The elastic atom-dimer scattering phase shift
         $\cot\delta_{AD}$ for $r^2\kappa^2=0,\pm 10^{-3},\pm 10^{-2}$
         as a function of the wave number $k/\kappa$.}
	\label{fig:cotdeltaAD}
\end{figure}

Next, we consider the effective range corrections to elastic atom-dimer 
scattering. We find scattering observables in general to be less sensitive
to the unphysical deep poles.
To obtain the scattering amplitude, we solve
Eq.~(\ref{eq:STM2}) for $E=\frac{3}{4m}k^2-E_2$ below the dimer breakup
threshold with the incoming particles on-shell. 
The elastic scattering phase shift
$\delta_{AD}(k)$ can then be obtained from the scattering amplitude
for $p=k$ using
\beq
\label{eq:tmatrix}
T\left(k,k;\frac{3}{4m}k^2-E_2\right)=\frac{3}{m}f_k
=\frac{3}{m}\frac{1}{\cot\delta_{AD}(k)-i}\,.
\eeq
In Fig.~\ref{fig:cotdeltaAD}, we show $\cot\delta_{AD}$ 
for different values of $r^2 \kappa^2$
as a function of the wave number $k$ up to the dimer breakup threshold
$k\approx 1.15\kappa$.
For small values of $k$, $\cot\delta_{AD}$ is almost linear
in $\ln k$ but at about half the breakup wave number the behavior
becomes more complicated. For $r^2\kappa^2< 0$, the non-universal 
corrections increase $\cot\delta_{AD}$ compared to 
the universal result while for $r^2\kappa^2 >0$ the behaviour
depends on the value of $r^2\kappa^2$.   

\begin{figure}
	\centerline{\includegraphics*[width=10cm]{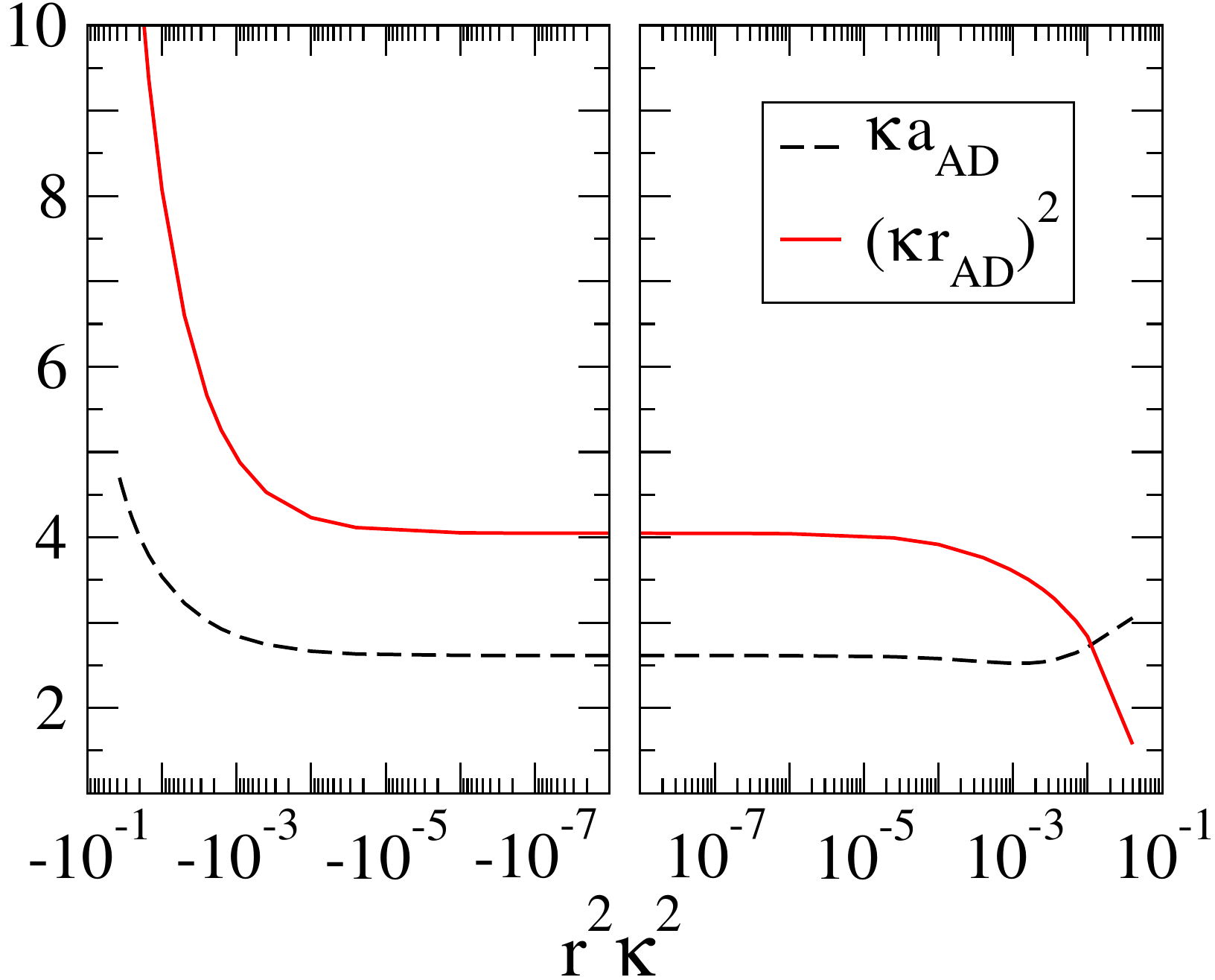}}
	\caption{(Color online) The atom-dimer scattering length 
          $\kappa a_{AD}$ and effective
          range $(\kappa r_{AD})^2$ as a function of the 
          two-body effective range $r^2 \kappa^{2}$.}
	\label{fig:aADrAD}
\end{figure}

The atom-dimer effective range parameters can be extracted from
our results by fitting the effective range expansion, Eq.~(\ref{eq:ere1}),
to $\cot\delta_{AD}$.
We have performed fits with different orders in the
expansion  and different truncations of the data 
sets to estimate the error in this extraction.
Our results for $\kappa a_{AD}$ and $(\kappa r_{AD})^2$ in
dependence of $r^2\kappa^2$ are summarized in Fig.~\ref{fig:aADrAD}.
The effective range  $r_{AD}^2$ comes out positive for all values
of $r^2\kappa^2$ considered.    
For $|r^2 \kappa^2|\lsim 10^{-4}$ the curves are nearly
symmetric around $r^2 \kappa^2=0$ and can be approximated by
\beq
\label{eq:aADpert}
\kappa a_{AD}=2.614(1)-4100(500)\, r^2\kappa^2+{\cal O}(r^4\kappa^4)\,.
\eeq
Similar to the bound state case, we find large coefficients in the 
perturbative expansion in $r^2 \kappa^2$.
For larger values of $r^2 \kappa^2$, the curves are not symmetric anymore.
In the unitary limit, we find the effective range parameters
$\kappa a_{AD}=2.614(1)$ and $(\kappa r_{AD})^2=4.0(2)$\,.
Converting to units of the scattering length
$a$, our results correspond to
$a_{AD}=2.328(1)\,a$ and $\ln(a_{AD}/a)=0.845(1)$.
These numbers agree well with  the value
$\ln(a_{AD}/a)=0.8451$ obtained by Kartavtsev et al.~\cite{Karta}
and are in qualitative agreement with the value $a_{AD}=2.95 a$ found
by Nielsen et al.~\cite{Nielsen99}. 
For $r^2\kappa^2$ of order 0.01, there are again substantial effective 
range effects. In particular, we find the values $\kappa a_{AD}=3.540(1)$ and
$(\kappa r_{AD})^2=7.4(2)$ for $r^2\kappa^2=-0.01$
and $\kappa a_{AD}=2.713(1)$ and $(\kappa r_{AD})^2=2.9(2)$
for $r^2\kappa^2=0.01$.

We have also calculated the atom-dimer scattering phase shifts
and effective range parameters using the fully perturbative treatment
discussed in Appendix~\ref{app:pert}. For sufficiently small effective
range, the two methods agree.

\subsection{Three-body recombination}
Finally, we consider three-body recombination into the shallow
dimer described by Eq.~(\ref{eq:Edimer}).
Cold atoms typically also have a large number of deep dimer states.
However, three-body recombination into the deep dimers is suppressed 
since the atoms have to approach to distances comparable to the 
size of the deep dimers. This process enters at the 
same order as short-range three-body interactions.
It could be calculated by introducing a complex three-body
parameter~\cite{Braaten:2004rn}.

The three-body recombination into the shallow dimer is strongly
influenced by the behavior of the full dimer propagator, Eq.~(\ref{eq:prop1}).
There are three limits in which the propagator
vanishes: $(i)$ $a\rightarrow \infty$, $(ii)$ $E\to 0$, and $(iii)$ 
$E\rightarrow \infty$. In these three limits, the three-body
recombination rate also vanishes.
For large but finite scattering length and finite energy, however,
three-body recombination can take place. 

\begin{figure}
\centerline{\includegraphics*[width=10cm]{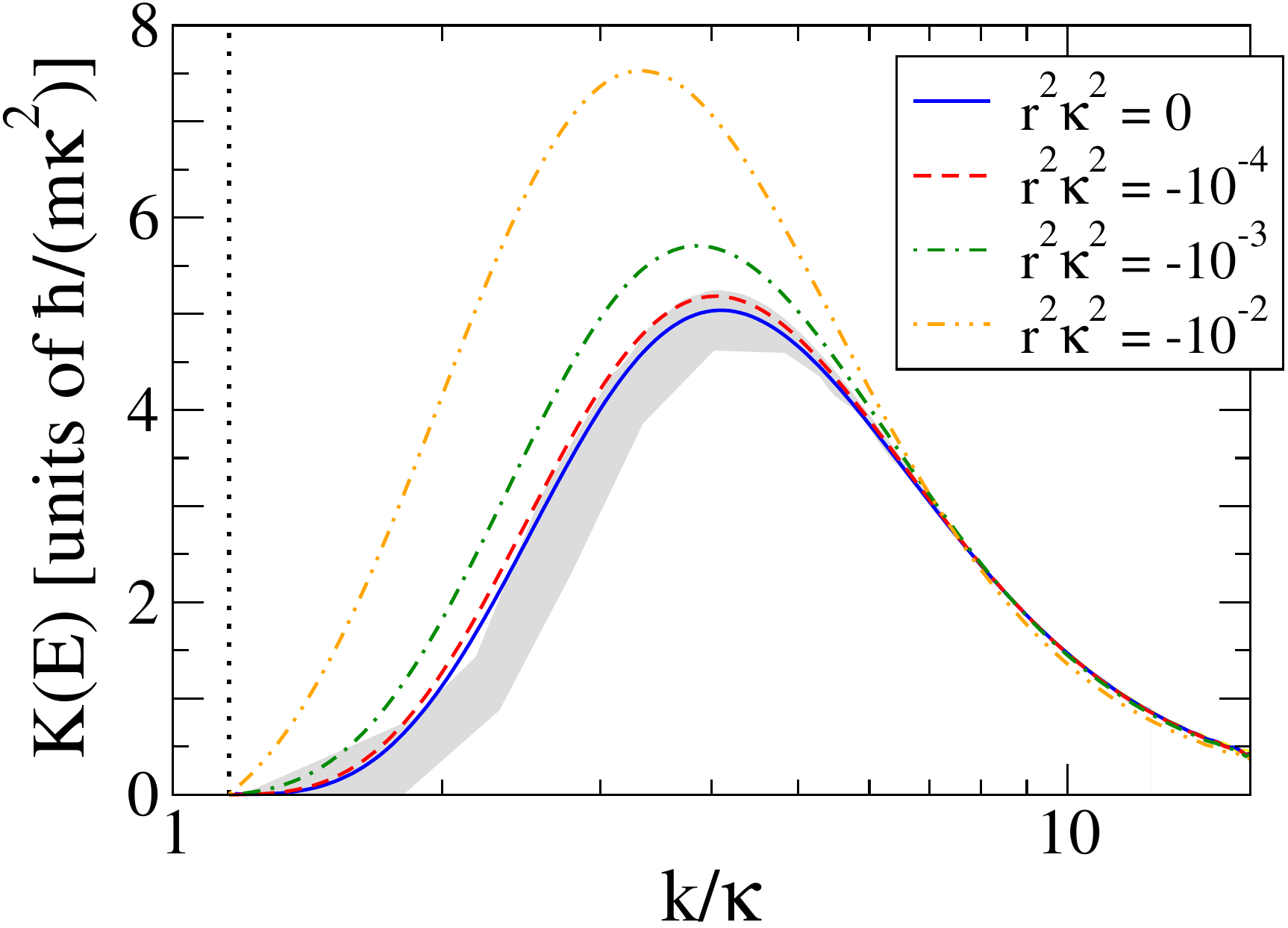}}
	\caption{(Color online) The energy-dependent 
         three-body recombination rate $K(E)$ in units of 
         $\hbar/(m\kappa^2)$
         as a function of the wave number $k/\kappa$. 
         The shaded band corresponds to $r^2\kappa^2=10^{-4}$ and is 
         derived with
         the help of cutoff variation as described in the text.}
	\label{fig:K}
\end{figure}

The rate can be conveniently calculated using
the inelastic atom-dimer scattering cross section~\cite{BHKP}. 
The integration measure of the three-body phase space in two dimensions 
using hyperspherical variables is given by
\beq
\label{eq:3bPS}
d^2p_1d^2p_2d^2p_3=m^2E\sin(2\alpha_3)dEd\alpha_3 d\varphi_{12}
d\varphi_{3,12}d^2p_{tot}\ .
\eeq
From this, the relation between the hyperangular average of the 
recombination rate at finite energy and the inelastic cross section in 
two dimensions is obtained as
\beq
\label{eq:K}
K(E)=\frac{36 \pi}{m^2 E}k\,
\sigma^{\rm (inel)}_{AD}(E)\, ,
\eeq
where $k=\sqrt{\frac{4m}{3} (E+E_2)}$.
The inelastic cross section in Eq.~(\ref{eq:K}) can be obtained by 
subtracting the elastic cross section
\beq
\label{eq:ela}
\sigma^{\rm (el)}_{AD}(E)=\frac{4}{k}|f_k|^2\, ,
\eeq
from the total cross section $\sigma^{\rm (tot)}_{AD}(E)$. 
The latter can be obtained from the optical
theorem which in $2d$ is given by~\cite{Adhikari1986}
\beq
\label{eq:opt}
\sigma^{\rm (tot)}_{AD}(E)=\frac{4}{k}{\rm Im} f_k(0)\, .
\eeq
Using the expression for the elastic scattering amplitude,
Eq.~(\ref{eq:tmatrix}), the three-body recombination rate $K(E)$
can be calculated from Eqs.~(\ref{eq:K}), (\ref{eq:ela}), and~(\ref{eq:opt}).
Our results for $K(E)$ as a function of the wave number $k$
are shown in Fig.~\ref{fig:K}. 
For positive $r^2 \kappa^2 $, we again make use of a cutoff variation in 
the range of 1/5 to 4/5 of the maximum allowed value.
The grey band gives our result for $r^2 \kappa^2 =10^{-4}$.
For larger values of $r^2 \kappa^2$, the width of the band increases.
As expected, the rate vanishes for large wave numbers and at threshold.
Around $k\approx 3\kappa-4\kappa$, there is a maximum in the recombination
rate. The position of the maximum is weakly dependent on the value
of the effective range.
It is governed by the behavior of the full dimer
propagator as a function of the energy, Eq.~(\ref{eq:prop1}). 
The maximum is not
related in a simple way to the energy of the universal three-body 
states, Eq.~(\ref{eq:3bdy2D}), but
our calculation establishes an implicit relation between the two.

\begin{figure}
\centerline{\includegraphics*[width=10cm]{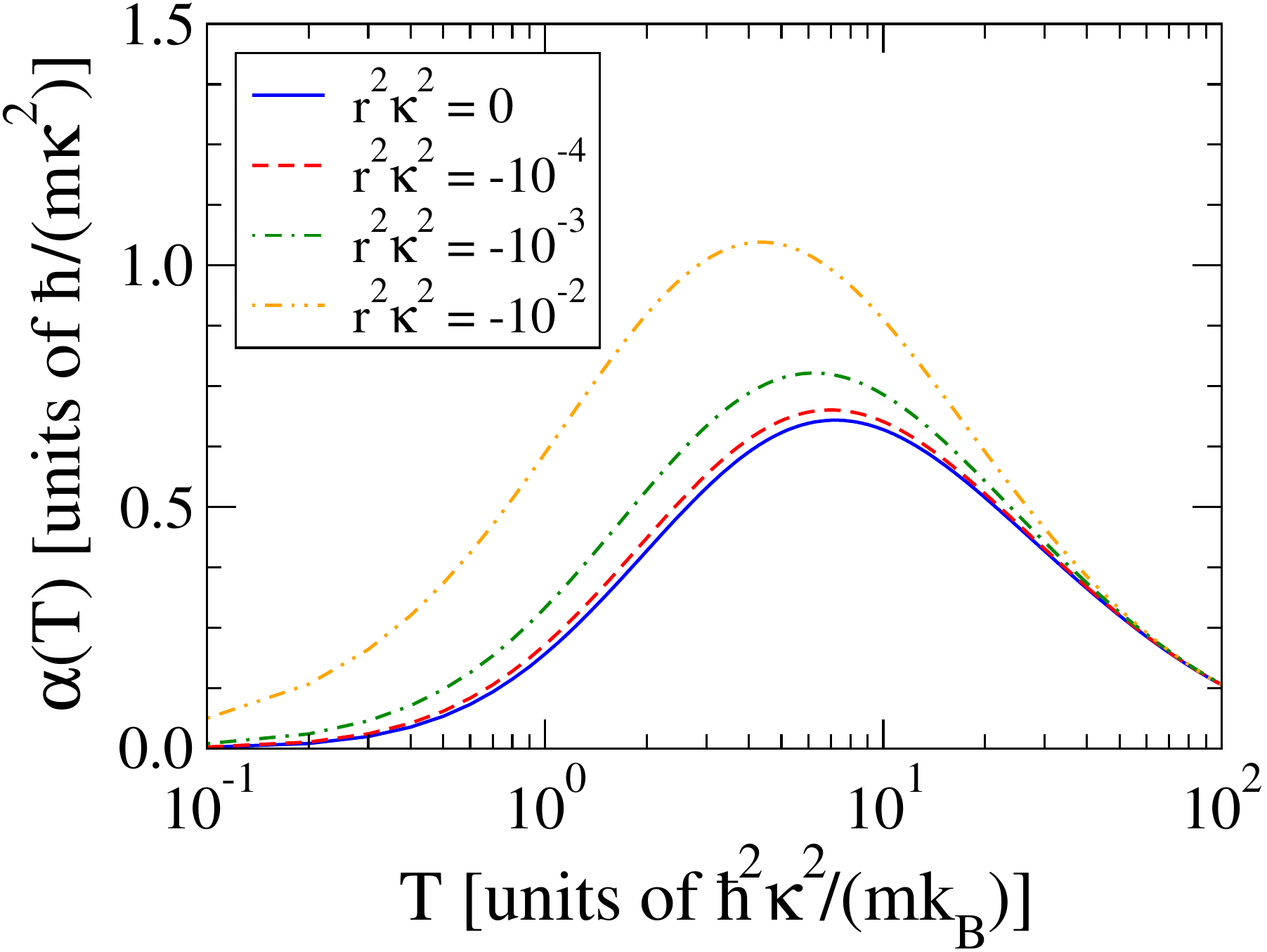}}
	\caption{(Color online) The three-body recombination rate $\alpha$ 
          in units of $\hbar/(m \kappa^2)$ in dependence of
          the temperature $T$ in units of $\hbar^2\kappa^2/(mk_B)$.}
	\label{fig:alpha}
\end{figure}

In experiments with cold atoms, one typically uses ensembles of atoms 
in thermal equilibrium.
The energy-dependent recombination rate $K(E)$ can be converted into
an energy averaged rate by performing a Boltzmann average
as described in Ref.~\cite{BHKP}. Taking into account the energy 
dependence of the three-body phase space, Eq.~(\ref{eq:3bPS}),
we have
\bqa
\alpha(T)&=&\frac{\int_0^\infty dE E e^{-E/(k_BT)}K(E)}{3!\int_0^\infty
dE E e^{-E/(k_BT)}}\nonumber\\ 
&=&\frac{1}{6k_B^2T^2}\int_0^\infty dE E e^{-E/(k_BT)}K(E)\, .
\eqa
Our results for $\alpha(T)$ for different values of the 
effective range are shown in  Fig.~\ref{fig:alpha}. We do not give
results for $r^2 \kappa^2 >0$ but the shaded band for $r^2 \kappa^2 =10^{-4}$ 
in Fig.~\ref{fig:K} translates into an error band around $r^2 \kappa^2 =0$ 
here as well.
The temperature dependent rate also has a maximum at temperatures
of the order of 5...7  times the dimer binding energy. The recombination 
rate at the maximum is very sensitive to the value of the effective range.
If the effective range is changed from zero to $r^2\kappa^2=-0.01$,
the rate at the maximum changes by a factor of two. For all values 
of the effective range considered, however, the recombination rate
at the maximum remains of order one in natural units $\hbar/(m\kappa^2)$.
This suggests that $2d$ Bose gases are stable enough
to observe universal few-body phenomena experimentally. 
The lifetime of a $2d$ Bose gas with large scattering length was
previously estimated by Pricoupenko and Olshanii~\cite{Pricoup07}.
The order of magnitude of our recombination rates is consistent with 
their results.

\section{Summary \& Conclusions}
In this work, we have investigated the 
three-body properties of identical bosons
close to the unitary limit in two spatial dimensions. 
Within an effective field theory for resonant interactions,
we have calculated the leading non-universal corrections 
which are due to the two-body effective range.
In particular, we have calculated the leading corrections
to the three-body binding energies, the atom-dimer scattering 
phase shift and effective range parameters, and the three-body 
recombination rate at finite energy.

We have compared our results to previous calculations in the 
unitary limit where available and generally found good agreement. 
Our calculations show a large sensitivity of three-body 
observables to the effective range. Significant effective range effects 
can be observed already for $|r^2\kappa^2|\gsim 10^{-4}-10^{-3}$. These
corrections are due to large coefficients in the perturbative 
expansion of observables in $r^2\kappa^2$ (cf.~Eqs.~(\ref{eq:E3pert}),
(\ref{eq:aADpert})). It would be interesting to understand the 
physics behind these large coefficients.
These coefficients could be reduced by an order of magnitude
by renormalizing to the three-body ground state energy $E_3^{(0)}$
instead of $E_2$~\cite{Birseprivate}, but they would still remain 
unnaturally large.
Our results suggest that the approach to the unitary limit in $2d$ 
three-body observables
is rather slow and effective range corrections play an important role
even close to the unitary limit. This is in agreement with the results 
of Blume who investigated the universal properties of 
$N$-body droplets using Lennard-Jones potentials and realistic Helium 
potentials~\cite{Blume2005}.

Our calculation of the three-body energies including the leading 
non-universal corrections
suggests that the bound states eventually cross the atom-dimer 
threshold as the effective range is made more negative. 
If this behavior holds true when higher orders are included,
it opens the possibility to observe three-body states in $2d$ 
through a variation of the $2d$ effective range.
The states would then appear as zero energy 
scattering resonances similar to Efimov states in $3d$.

Our results are directly applicable to two-dimensional Bose gases 
with large scattering length and
imply that effective range effects must be under control
in experiments on universal properties of $2d$ Bose gases.
Effective field theory provides a powerful tool to calculate
these corrections and our study provides the first step towards
accurate calculations of these effects. 

On the experimental side, there has been some progress in the study of
universal properties of $2d$ Bose gases. For example,
Chin and coworkers have recently studied scale invariance and critical
behavior near a Berezinsky-Kosterlitz-Thouless phase transition in $2d$ and observed universal behavior
of the thermodynamic functions~\cite{Chin2010}.
They have also attempted to show how the Efimov
resonance in three-body recombination shifts when the system is tuned
towards a dimensionality of two by increasing one of the trapping
frequencies in a $3d$ experiment~\cite{Chin2009}. 

Interesting few-body properties of $2d$ systems include 
universal $N$-body states and a geometric spectrum of 
$N$-body ground states~\cite{Son}. Moreover, Nishida and Tan have shown 
that  a two-species Fermi gas in which one species is confined in 
$2d$ or $1d$ while the other is free in the three-dimensional space
is stable against the Efimov effect and has universal 
properties~\cite{Nishida:2008kr}. More complicated multispecies Fermi gases 
with similar properties are possible as well.
They also showed that a purely $S$-wave resonance in $3d$ can induce
higher partial wave resonances in mixed dimensions and
pointed out that some of the resonances observed in a recent
experiment by the Florence group~\cite{Florence10} can be interpreted as
a $P$-wave resonance in mixed $2d$-$3d$ dimensions~\cite{Nishida:2010mw}.
Thus, future experiments considering few-body phenomena
in lower-dimensional ultracold gases will be very interesting.
Our calculation of the leading non-universal corrections 
provides a basis for the interpretation of such experiments.

\begin{acknowledgments}
We thank Y.~Nishida, L.~Pricoupenko, and D.\ Phillips for discussions and 
D.~Blume for providing her data. We are also grateful to the INT in 
Seattle for hospitality during the program 
\lq\lq Simulations and Symmetries: Cold Atoms, QCD, and
Few-hadron Systems'' where part of this work was done. 
K.H.\ was supported by the ``Studienstiftung des deutschen Volkes''
and by the Bonn-Cologne Graduate School of Physics and Astronomy.
\end{acknowledgments}

\begin{appendix}

\section{Perturbative treatment}
\label{app:pert}

It is also possible to include the effective range in a fully
perturbative way. In order to achieve this, we expand 
the boson-dimer scattering amplitude into a leading order piece
$T_0(p,k;E)$ which satisfies Eq.~(\ref{eq:STM2}) with $r^2\equiv 0$
and a correction $T_2(p,k;E)$ of order $r^2$:
\beq
T(p,k;E)=T_0(p,k;E)+T_2(p,k;E)+\ldots\,.
\label{eq:expT}
\eeq
Next, we insert this expansion~(\ref{eq:expT}) into Eq.~(\ref{eq:STM2}),
expand the $r^2$ dependent terms, and collect all terms of order 
$r^2$ in order to obtain an equation for $T_2(p,k;E)$.  
The on-shell scattering amplitude at next-to-leading order can then
be written as an integral over the leading order amplitude and 
we finally obtain:
\bqa
\label{eq:Tpert}
T(k,k;E)&=&\left(1+\frac{r^2 \kappa^2}{2}\right)\,T_0(k,k;E)\nonumber \\
&-&\frac{mr^2}{4\pi\kappa^2} \int_0^\infty dq\, q\, 
    \left(\kappa^2+mE-\frac{3}{4}q^2\right)\,
    \left[\frac{T_0(k,q;E)}{\ln\left((\frac{3}{4}q^2-mE-i\epsilon)/\kappa^2
          \right)}\right]^2\,,
\eqa
where $E=3k^2/(4m)-E_2$. The scattering phase shift can be extracted
using Eq.~(\ref{eq:tmatrix}) as before.

\end{appendix}


\end{document}